\documentclass[12pt,cite,epsf,epsfig]{article}
\usepackage{epsfig}

\begin{document}

\author{Sanjeev Kumar}
\title{Unitarity constraints on trimaximal mixing}

\date{\textit{Department of Physics and Astrophysics, University of Delhi,\\ Delhi -110005, INDIA.}\\
email: sanjeev3kumar@gmail.com\\
}

\maketitle

\begin{abstract}
When the neutrino mass eigenstate $\nu_2$ is trimaximally mixed,
the mixing matrix is called trimaximal. The middle column of trimaximal
mixing matrix is identical to tri-bimaximal mixing and the other two
columns are subject to unitarity constraints. This corresponds to a
mixing matrix with four independent parameters in the most general 
case. Apart from the two Majorana phases, the mixing matrix has only
one free parameter in the CP conserving limit. Trimaximality results
into interesting interplay between mixing angles and CP violation. A
notion of maximal CP violation naturally emerges here: CP violation is
maximal for maximal 2-3 mixing. Similarly, there is a natural 
constraint on the deviation from maximal 2-3 mixing which takes its
maximal value in the CP conserving limit.

\end{abstract}

\section{Introduction}

The historical discovery of neutrino oscillations
constitutes the only indication of physics beyond the
standard model. In the absence of a basic understanding of fermion
masses and mixings, it is natural to seek simplified patterns in
the mixing matrix which are consistent with the
present experimental knowledge and provide guidance for the future 
experimental searches. Such phenomenological patterns are, at best,
first approximations to the experimental data and the next step is 
to study deviations from these patterns. 

A successful phenomenological ansatz for neutrino mixing matrix which 
is consistent with the present neutrino data \cite{pdg} was 
proposed by Harrison, Perkins and Scott \cite{tbm} and is given by
\begin{equation}
U=\left( \begin{array}{ccc}
\sqrt{\frac{2}{3}} & \frac{1}{\sqrt{3}} & 0 \\ 
-\frac{1}{\sqrt{6}} & \frac{1}{\sqrt{3}} & \frac{1}{\sqrt{2}} \\ 
-\frac{1}{\sqrt{6}} & \frac{1}{\sqrt{3}} & -\frac{1}{\sqrt{2}}
\end{array} \right).
\end{equation}
In this mixing scheme, the second neutrino mass eigenstates $\nu_2$ 
has trimaximal character as it arises from maximal mixing of the three 
flavor eigenstates ($\nu_{e}$, $\nu_{\mu}$ and $\nu_{\tau}$). Moreover, 
the third mass eigenstate $\nu_3$ has bimaximal character as it 
results from maximal mixing of two flavor eigenstates \textit{viz.} 
$\nu_{\mu}$ and $\nu_{\tau}$. Hence, this mixing scheme is called 
tri-bimaximal (TBM) mixing and gives vanishing $\theta_{13}$ and 
maximal $\theta_{23}$. The mixing angle $\theta_{13}$ vanishes in TBM 
mixing because of bimaximal character of $\nu_3$, a 
feature needs to be abandoned in order to allow for a non-vanishing 
$\theta_{13}$ and, hence, the possibility of CP violation. However, 
there is no need to abandon the trimaximal character of the $\nu_2$ in 
this generalization of TBM mixing. Such a mixing matrix, called 
trimaximal (TM) \cite{lam-tm-magic}, will be of the form
\begin{equation}
U=\left( \begin{array}{ccc}
U_{11} & \frac{1}{\sqrt{3}} & U_{13} \\ 
U_{21} & \frac{1}{\sqrt{3}} & U_{23} \\ 
U_{31} & \frac{1}{\sqrt{3}} & U_{33}
\end{array} \right)
\end{equation}
where the first and third columns are subject solely to unitarity 
constraints and the middle column has trimaximal character. The 
definition of trimaximality adopted here requires
that magnitudes and phases of all the three elements in the 
middle column of the mixing matrix are equal. The common phase of
the middle column can be rephased away by phase-redefinition of the
lepton fields. 

The relationship between TBM mixing and TM mixing needs also to
be understood from the standpoint of the symmetry 
groups associated with these mixing schemes. The TM mixing follows
from a neutrino mass matrix which can be parametrized as
\begin{equation}
M=\left( \begin{array}{ccc}
a+2d & c-d & b-d\\ 
c-d & b+2d & a-d \\ 
b-d& a-d & c+2d
\end{array} \right)
\end{equation}
in terms of four complex parameters $a$, $b$, $c$ and $d$ 
having $A_4$ symmetry in the flavor basis where the charged
lepton mass matrix is diagonal \cite{lam-a4}. The matrix is called 
magic mass matrix since its rows and columns all add up to $a+b+c$ and 
hence one of its eigenvectors is $\left(1/\sqrt3~~1/\sqrt3~~1/\sqrt3\right)^T$ within an overall
multiplicative phase \cite{lam-tm-magic}. The above mass matrix 
becomes $\mu$-$\tau$ symmetric as well if the number of free 
parameters are reduced to three by demanding $b=c$. The mixing 
matrix become TBM and the symmetry group enlarges to $S_4$ after 
making the above fine tuning. Thus, the neutrino mass matrix having
$A_4$ symmetry and bringing about TM mixing contains one more complex 
parameter than the mass matrix having  $S_4$ symmetry and producing 
TBM mixing which can be used to accommodate a non-vanishing $\theta_{13}$
and a Dirac-type CP-violating phase $\delta$. This additional 
parameter needs to be fine tuned to recover the TBM mixing in a
mass model based upon $A_4$ symmetry. The neutrino mixing matrix 
will be TM in absence of any such fine tuning of
parameters \cite{lam-a4}. It becomes imperative to analyse the 
phenomenology of TM mixing scheme in this context. 

The CP violating structure in trimaximal mixing is an 
interesting feature to study because of the mathematical 
restriction that one of the eigenvector of the mass matrix be 
$\left(1~~1~~1\right)^T$ which severely
restricts the elements of the matrix to coordinate in a 
way so that they produce a real eigenvector. This coordination results 
into a unique correlation between CP violation and the
mixing angles.

A notion of maximal CP violation automatically emerges when unitarity 
constraints are combined with trimaximality. The Jarlskog rephasing 
invariant $J$ \cite{jarlskog} takes its maximal value $\frac{1}{2\sqrt{3}}|U_{11}||
U_{13}|$ when $\theta_{23}$ is maximal. Similarly, the deviation from 
maximal 2-3 mixing is proportional to $\theta_{13}$ and is naturally 
restricted to lie within the present experimental range by unitarity 
for values of $\theta_{13}$ within its experimental bounds \cite{pdg}.
This deviation is zero when CP violation is maximal and is maximal 
when there is no CP violation.

\section{Mixing angles and CP violating phases}
Trimaximality, as defined in Eq. (2), equates magnitudes as well as 
phases of the three elements of the middle column of $U$. Although, 
the equality of the magnitudes is preserved after the phase 
redefinitions of the lepton fields, same is not true for the equality 
of the phases. Hence, the definition of trimaximal mixing adopted here
is not rephasing invariant. This is quite natural since a symmetry
may prefer a basis, called symmetry basis, for its definition. 
So, it becomes necessary to find the consequences of TM mixing in 
terms of rephasing invariant quantities. The mixing angles and
CP violating phases must be deduced from these rephasing invariants 
only.

There are four independent CP-even quadratic invariants \cite{rephasing_inv} which can 
conveniently be chosen as $U^{*}_{11}U_{11}$, $U^{*}_{13}U_{13}$, $U^{*}_{21}U_{21}$ and $U^{*}_{23}U_{23}$ and three independent CP-odd quartic invariants \cite{rephasing_inv}
\begin{equation}
J=Im(U_{11}U^*_{12}U^*_{21}U_{22}),
\end{equation}
\begin{equation}
I_1=Im[(U^*_{11}U_{12})^2]
\end{equation}
and
\begin{equation}
I_2=Im[(U^*_{11}U_{13})^2].
\end{equation}
The Jarlskog rephasing invariant $J$ \cite{jarlskog} is relevant for CP violation in 
lepton number conserving process like neutrino oscillations and 
$I_1$ and $I_2$ are relevant for CP violation in lepton number 
violating processes like neutrinoless double beta decay.

The consequences of unitarity for TM mixing will be worked out in terms 
of the above rephasing invariant quantities \textit{viz.} $|U_{11}|$, 
$|U_{13}|$, $|U_{21}|$, $|U_{23}|$, $J$, $I_1$ and $I_2$. However, it 
will be rewarding to express the phenomenological implications of 
trimaximality in term of the usual parameters $\theta_{12}$, 
$\theta_{23}$, $\theta_{13}$, $\alpha$, $\beta$ and $\delta$ of a 
widely used parametrisation \cite{pdg,pdg-para}
\begin{equation}
U= \left(
\begin{array}{ccc}
c_{13}c_{12} & c_{13}s_{12} & s_{13}e^{-i\delta} \\
-c_{23}s_{12}-c_{12}s_{23}s_{13}e^{i\delta} &
c_{12}c_{23}-s_{12}s_{23}s_{13}e^{i\delta} & c_{13}s_{23}\\
s_{12}s_{23}-c_{12}c_{23}s_{13}e^{i\delta} &
-c_{12}s_{23}-c_{23}s_{12}s_{13}e^{i\delta} & c_{13}c_{23}
\end{array}
\right)
\left(
\begin{array}{ccc}
1 & 0 & 0 \\ 
0 & e^{i \alpha} & 0 \\ 
0 & 0 & e^{i (\beta+\delta)}
\end{array}
\right)
\end{equation}
where $s_{ij}=\sin\theta_{ij}$ and $c_{ij}=\cos\theta_{ij}$. This will be achieved through the relations $|U_{11}|=c_{13}c_{12}$,  $|U_{13}|=s_{13}$, $|U_{23}|=c_{13}s_{23}$, 
\begin{equation}
I_1=c^4_{13}c^2_{12}s^2_{12}\sin 2 \alpha,
\end{equation}
\begin{equation}
I_2=c^2_{13}c^2_{12}s^2_{13}\sin 2 \beta
\end{equation}
and
\begin{equation}
J=c^2_{13}c_{23}c_{12}s_{12}s_{23}s_{13}\sin \delta.
\end{equation}
However, some of the relationships will be more apparent when 
written in terms of rephasing invariant quantities themselves.

Like $\mu-\tau$ symmetry and TBM mixing, the TM mixing is 
mass-independent texture putting no constraints on neutrino masses and 
Majorana phases $\alpha$ and $\beta$ \cite{lam-mit}. Hence, 
trimaximality will have no physical consequences for the Majorana 
phases. 

\section{Strong and weak trimaximality}
A naive way to work out the consequences of unitarity for trimaximal
mixing would have been to compare the magnitudes of the elements of 
trimaximal mixing matrix of Eq. (2) with the unitary mixing matrix of 
Eq. (7). This would mean that $|U_{21}|$, 
$|U_{22}|$ and $|U_{23}|$ in Eq. (7) are all equal to 
$\frac{1}{\sqrt{3}}$ with no restriction on their phases. In this 
version of TM mixing, trimaximality is imposed 
only in a weak sense as the elements of the middle column of $U$ in 
Eq. (7) have unequal phases for any non zero value of $\delta$ and 
$\theta_{13}$. This weak form of trimaximality has been already 
analysed in the literature \cite{tm-albright} for its phenomenological 
consequences. If a similar analysis to is to 
be performed for the trimaximality defined by Eq. (2), called strong 
trimaximality to distinguish it from weak trimaximality, which 
demands that magnitudes as well as the phases of $U_{21}$, $U_{22}$ 
and $U_{23}$ are equal, one is required to compare 
the trimaximal $U$ of Eq. (2) with the unitary $U$ of Eq. (7) only after 
making its middle column completely real by making a phase rotation on 
the lepton fields. This process of making a 
basis transformation on a mixing matrix to the symmetry basis where 
the TM mixing is defined is quite natural if the definition of
symmetry is not rephasing invariant and prefers some basis.
As stated already, the strong trimaximality is not a rephasing 
invariant concept unlike weak trimaximality which is a rephasing 
invariant notion. However, instead of adapting a unitary matrix by 
phase rotation to match it with trimaximal mixing as outlined above, 
the present work attempts to obtain an unitary parametrisation which 
is trimaximal in the strong sense from the very outset.

The strong trimaximility implies that the mass eigenstate $\nu_{2}$ 
is a democratic and coherent combination of the three
flavor eigenstates whereas the week trimaximality relaxes the 
condition of coherence allowing different flavor eigenstates
to mix with equal amplitudes but mismatching phases. However, the distinction between 
strong and weak trimaximality is not merely
conceptual, it is also phenomenological in nature having distinct
predictions for CP violation. The weak trimaximality is less 
interesting as it doesn't fully utilise the phase restrictions on the 
mass matrix coming from $A_4$ symmetry. 

A mixing matrix which is trimaximal in the strong sense was first 
proposed and studied phenomenologically by Bjorken, Harrison and Scott
(BHS) \cite{bhs} in which the only free complex parameter was 
$U_{13}$. A neutrino mass matrix having magic symmetry in the basis of 
a diagonal charged lepton mass matrix was prescribed by 
Lam \cite{lam-magic} which gives rise to TM mixing in BHS 
parametrization. In another attempt for the minimal modification to 
the TBM mixing, He and Zee \cite{minimal-a4} obtained a mixing
matrix, which is also trimaximal in strong sense, containing only one 
free complex parameter. The neutrino mass matrix in a model proposed 
by Grimus and Lavoura \cite{tm model} is magic and hence gives rise 
to TM mixing in the strong sense. However, the mass matrix contains 
three complex parameters and therefore is more restrictive than
a most general magic mass matrix [Eq. (2)] containing four complex 
parameters. Similarly, the resulting TM mixing matrix also contains
lesser number of parameters than the most general case. 
All the above forms of TM mixing studied phenomenologically 
in the literature are more constrained than the most general
form given by Eq. (2) and can be expressed as the product of 
the TBM mixing matrix of Eq. (1) by a generalized 1-3 rotation of the 
type
\[
\left( \begin{array}{ccc}
\cos \theta & 0 & \sin \theta ~e^{-i \phi}\\ 
0 & 1 & 0 \\ 
-\sin \theta ~e^{i \phi}& 0 & \cos\theta
\end{array} \right)
\]
from the right. However, the resulting mixing matrix 
is not the most general TM mixing matrix.

\section{Unitary constraints}

The TM mixing matrix of Eq. (2) can be written as
\begin{equation}
U=\left( \begin{array}{ccc}
U_{11} & \frac{1}{\sqrt{3}} & U_{13} \\ 
U_{21} & \frac{1}{\sqrt{3}} & U_{23} \\ 
-U_{11}-U_{21} & \frac{1}{\sqrt{3}} & -U_{13}-U_{23}
\end{array} \right)
\end{equation}
by straightforwardly eliminating two elements \textit{viz.} 
$U_{31}$ and $U_{33}$ using the two orthogonality conditions 
for the three columns. The remaining 4 elements are 
further constrained by the following seven constraints coming from the 
unitarity condition
$UU^{\dagger}=U^{\dagger}U=1$:
\begin{equation}
|U_{11}|^2+|U_{13}|^2=\frac{2}{3},
\end{equation}
\begin{equation}
|U_{21}|^2+|U_{23}|^2=\frac{2}{3},
\end{equation}
\begin{equation}
|U_{11}|^2+|U_{21}|^2+Re{U^*_{11}U_{21}}=\frac{1}{2},
\end{equation}
\begin{equation}
|U_{13}|^2+|U_{23}|^2+Re{U^*_{13}U_{23}}=\frac{1}{2},
\end{equation}
\begin{equation}
Im{(U^*_{11}U_{21}})+Im({U^*_{13}U_{23}})=0,
\end{equation}
\begin{equation}
2Re{(U_{11}U^*_{13}})+Re{(U_{21}U^*_{13}})+Re{(U_{11}U^*_{23}})
+2Re{(U_{21}U^*_{23}})=0
\end{equation}
and
\begin{equation}
2Im{(U_{11}U^*_{13}})+Im{(U_{21}U^*_{13}})+Im{(U_{11}U^*_{23}})
+2Im{(U_{21}U^*_{23}})=0.
\end{equation}
Eq. (11) is a unitary parametrisation of TM mixing subject to the 
above seven unitarity constraints. All the parametrizations of TM 
mixing proposed in the literature \cite{bhs,minimal-a4,tm model} 
are special cases of the above parametrization and satisfy these 
seven constraints. However, they are more constrained than the 
general parametrization obtained here. A special property of the
TM mixing matrix of Eq. (11), also shared by its earlier 
parametrizations \cite{bhs,minimal-a4,tm model}, is that its first 
and third columns add up to zero. 

The first four of the seven unitarity constraints can be seen 
as normalization conditions and the last three as orthogonalization 
conditions. The three orthogonalization relations can be solved
to get three real variables, say, $Re U_{23}$, $Im U_{23}$ and 
$Im U_{21}$ which can be substituted back into the four 
normalization relations. This eliminates these three parameters. Two 
classes of solutions are obtained on solving the four normalisation 
relations in the remaining five parameters: one is CP conserving while 
the other is CP violating.

\section{The CP conserving solution}

The first solution has three independent parameters which can be 
chosen as $ReU_{11}$, $ReU_{13}$ and $ImU_{13}$. The imaginary part of 
$U_{11}$ can be calculated from the equation
\begin{equation}
|U_{11}|=\sqrt{2/3-|U_{13}|^2}.
\end{equation}
The elements $U_{21}$ and $U_{23}$ are given by
\begin{equation}
U_{21}=\frac{U_{11}}{2}\left(-1\pm\frac{\sqrt{3}|U_{13}|}{|U_{11}|}\right)
\end{equation}
and
\begin{equation}
U_{23}=\frac{U_{13}}{2}\left(-1\mp \frac{\sqrt{3}|U_{11}|}{|U_{13}|}\right).
\end{equation}
The resulting mixing matrix has no Dirac type CP violation
since first two elements in either first or third column have 
identical phases which results into a vanishing $J$. There are no
constraints on the phases of $U_{11}$ and $U_{13}$ and they are not 
necessary to describe the phenomenon of neutrino oscillations. 
However, the CP-odd rephasing invariants $I_1$ and $I_2$ do depend 
upon these phases and are given by 
\begin{equation}
I_1=-\frac{2}{3}ReU_{11} ImU_{11}
\end{equation}
and
\begin{equation}
I_2=Re(U_{11}^2)Im(U_{13}^2)-Im(U_{11}^2)Re(U_{13}^2).
\end{equation}

This solution is called $CP$ conserving as there is no Dirac type
CP violation here. Trimaximality is a mass-independent texture and it 
puts no constraints on neutrino masses and Majorana phases 
\cite{lam-mit}. Hence, a special case with vanishing Majorana phases 
will have real $U_{11}$ and $U_{13}$ parametrizable by a single 
parameter, say $\theta$. The mixing matrix will be completely
real in this special case and is given by
\begin{equation}
U=\left( \begin{array}{ccc}
\sqrt{\frac{2}{3}}c & \frac{1}{\sqrt{3}} & \sqrt{\frac{2}{3}} s \\ 
-\frac{c}{\sqrt{6}}\pm\frac{s}{\sqrt{2}} & \frac{1}{\sqrt{3}} & -\frac{s}{\sqrt{6}}\mp\frac{c}{\sqrt{2}} \\ 
-\frac{c}{\sqrt{6}}\mp\frac{s}{\sqrt{2}} & \frac{1}{\sqrt{3}} & -\frac{s}{\sqrt{6}}\pm\frac{c}{\sqrt{2}} \\ 
\end{array} \right)
\end{equation}
where $c=\cos\theta$, $s=\sin\theta$ and the parameter $\theta$  can be 
related to the mixing angle $\theta_{13}$ through the relation $\sin\theta_{13}=\sqrt{\frac{2}{3}}\sin\theta$.
This mixing matrix coincides with the mixing matrix obtained in a 
lepton mass model proposed by Grimus,  Lavoura and Singraber 
\cite{tm-cp conserving} for the first sign choice.

\section{The CP violating solution}

The second solution has four independent parameters which can be 
chosen as $ReU_{11}$, $ReU_{13}$, $ReU_{21}$ and $ImU_{13}$. The 
imaginary parts of $U_{11}$ and $U_{21}$ are calculated from the 
equations
\begin{equation}
|U_{11}|=\sqrt{2/3-|U_{13}|^2}
\end{equation}
and
\begin{equation}
Im U_{21}=-\frac{1}{2}\left(Im U_{11}+D\right)
\end{equation}
where the real parameter $D$ is given by
\begin{equation}
D^2=3 |U_{13}|^2-\left[Re(U_{11}+2U_{21})\right]^2
\end{equation}
which can either be positive or negative. The element $U_{23}$ is given by
\begin{equation}
U_{23}=-\frac{1}{2}\left[U_{13}+
\frac{U_{11}^*}{U_{13}^*} \left\{
Re(U_{11}+2U_{21})-iD
\right\}
\right].
\end{equation}
The Jarlskog invariant $J$ is given by
\begin{equation}
J=\frac{1}{6}\left(Re U_{11}Im U_{11}+2 Re U_{21}Im U_{11}+D Re U_{11}\right).
\end{equation}

The solution is called CP violating solution since $J$ is non-zero
here. The CP conserving solution can be recovered from this solution 
in the limiting case where $J=0$. $J$ vanishes for the two values of 
$U_{23}$ given by Eq. (21) which are also the two limiting values 
within which $U_{23}$ varies in the CP violating solution [Eq. (28)]. 
Thus, the CP conserving solution for $U_{23}$ defines the limiting 
region in which $U_{23}$ varies in the CP violating solution.

A notion of maximal CP violation naturally emerges here. 
It can be seen from the Eqs. (28) and (29) that $J$ is maximal when 
$\theta_{23}$ is maximal and the maximal value is given by
\begin{equation}
J^2=\frac{1}{12}|U_{11}|^2|U_{13}|^2.
\end{equation}
It should be noted that the maximal CP violation does not mean that 
the CP violating phase $\delta$ will also be maximum when $J$ and 
$\theta_{23}$ are maximal. The maximal value of $J$ obtained
here coincides with the intrinsic CP violation in the mass model by He 
and Zee \cite{minimal-a4}. This is because the mixing matrix in the 
model is a special case of the most general unitary TM 
mixing matrix considered here.

\section{Numerical analysis and results}

The relationship between $\theta_{12}$ and $\theta_{13}$ 
given by Eq. (19) or (25) has been depicted in Fig. 1 and is 
identical for both the solutions. The relationship is well known 
and has been reproduced here to show the values of $\theta_{12}$ 
and $\theta_{13}$ which will enter into the numerical calculations 
through $U_{11}$ and $U_{13}$. The value
of $\theta_{12}$ is constrained automatically to be well below 
its $3 \sigma$ range when $\theta_{13}$ is varied 
upto its $3\sigma$ upper bound \cite{pdg}.

Both the CP conserving and CP violating solutions have identical 
predictions for the $ee$ element of the neutrino mass matrix 
\begin{equation}
M_{ee}=m_1 U_{11}^2+\frac{m_2}{3}+m_3 U_{13}^2
\end{equation}  
whose magnitude, also called effective Majorana mass, will be 
observable in neutrinoless double beta day experiments. Varying the 
lowest neutrino mass over a reasonable range ($10^{-3}~eV$-$1~eV$), 
calculating other mass eigenvalues from the squared-mass differences $\Delta m^2_{12}$ and $\Delta m^2_{12}$ (Table 1) sampled using normal 
distribution and using the values of $\theta_{12}$ and 
$\theta_{13}$ sampled in Fig. 1, the predictions for $|M_{ee}|$ 
have been shown in Fig. 2 for normal and inverted orderings of 
neutrino masses. This graph matches with usual plots of $|M_{ee}|$ 
as the function of lowest neutrino mass and has been reproduced here 
merely to check correctness of the numerical analysis.

The variation of $|U_{23}|$ with $|U_{13}|$ for the CP conserving
solution can be calculated directly from Eq. (21) as
\begin{equation}
|U_{23}|=\frac{1}{2}\left( \sqrt{3}|U_{11}|\pm |U_{13}|\right)
\end{equation}
and has been plotted in Fig. 3. It is noted that $|U_{23}|$ depends 
only on $|U_{13}|$ for the CP conserving solution. However, $U_{23}$ 
depends on $ReU_{11}$, $ReU_{13}$, $ImU_{13}$ and $ReU_{21}$ in the 
CP violating 
solution. Varying these free parameters randomly in their allowed 
ranges, the variation of $|U_{23}|$ with $|U_{13}|$ 
for the CP violating solution can be calculated from Eq. (28) and has 
been shown in Fig. 4. One can easily see that the CP conserving 
solution depicted in Fig. 3 sets the upper and lower limits within 
which the CP violating solution shown in Fig. 4 can vary. The same 
fact is true for the variation between $\theta_{23}$ and  
$\theta_{13}$ depicted in Fig. 5 for CP conserving solution and in 
Fig. 6 for CP violating solution. The 
two branches of CP conserving solution for $\theta_{23}$ can  
be viewed as the bounds within which $\theta_{23}$ varies in 
the CP violating solution. It can be easily seen that $\theta_{23}$ 
can be maximal even for non-zero $\theta_{13}$ in trimaximal mixing. 
The allowed span of $\theta_{23}$ around the maximal value increases
with $\theta_{13}$. When $\theta_{13}$ reaches its $3\sigma$
upper bound, $\theta_{23}$ also spans its $3\sigma$ allowed range  \cite{pdg}. The span within which $\theta_{23}$ is allowed to vary
increases almost linearly with $\theta_{13}$ showing a non-linear 
behaviour only at very large values of $\theta_{13}$ which has not been 
depicted in Figures 5 and 6.  The 
deviation from maximal 2-3 mixing is zero when CP violation is maximal 
and it increases as CP violation decreases and, finally, there is no 
CP violation at the point where the deviation from maximality 
takes its maximal value allowed in TM mixing.

For the CP violating solution, $J$, $I_1$ and $I_2$ are calculated 
from Eqs. (22), (23) and (29) by varying the four free parameters 
$ReU_{11}$, $ReU_{13}$, $ImU_{13}$ and $ReU_{21}$ in their full ranges 
allowed by experimental as well as the unitarity constraints. The 
relationship between $J$ with $|U_{23}|$ for $|U_{13}|=0$, $0.1$ and 
$0.2$ has been depicted in Fig. 7. The central point is the TBM 
solution for which both $J$ and $U_{13}$ are zero and $|U_{23}|
=\frac{1}{\sqrt{2}}$. However, there are values of $|U_{23}|$ other 
than the TBM value for which $J$ is zero [Eq. (32)] and there are 
values of $J$ other than zero for which $|U_{23}|$ takes its TBM value 
[Eq. (30)]. The relationship between $U_{13}$, $U_{23}$ and $J$ shown
in Fig. (7) is unique to trimaximal mixing not found in other mixing
schemes deviating from TBM mixing \cite{tm-albright} (and refernces 
therein) and subject to verification in future neutrino oscillation 
experiments.

The relationship between $\delta$ and $\theta_{23}$ for 
$\theta_{13}=4$, $8$ and $12$ degrees has been depicted 
in Fig. 8. It is apparent from this figure that $\delta$ does not
attain its maximum value at $\theta_{23}=\frac{\pi}{4}$. So, the
point of maximal CP violation is not the point at which $\delta$
is maximum. The maximum of $\delta$ occurs when $\theta_{23}$ is 
slightly below the maximal value. Furthermore, the span in which 
$\theta_{23}$ is allowed to vary scales up with $\theta_{13}$ 
(see Fig. 6 as well) but the allowed range of $\delta$ is almost 
unaffected by variation in $\theta_{13}$ showing a marginal increase 
of about $10$ degrees with increase in $\theta_{13}$. The 
relationship between CP violation and deviation from maximality 
can be visualized in the most apparent way
in Fig. 9 which plots $s^2_{23}$ as a function of $J$ for 
$\theta_{13}=4$, $8$ and $12$ degrees. For a given
$\theta_{13}$, the allowed values of $J$ and $s_{23}^2$ form
an ellipse around the TBM point 
$\left(J=0,s_{23}^2=\frac{1}{2} \right)$. The semi-major axis of 
the ellipse is given by the maximal value of $J$ [Eq. (30)] and the 
semi-minor axis of the ellipse is defined by the CP conserving values 
of $s^2_{23}$ [Eq. (32)]. 

For completeness, the variation of $J$ 
and $I_2$ with $U_{13}$ has been depicted in Fig. 10. There are
no constraints on $I_1$. Since the phases of $U_{11}$ and $U_{13}$ are
not constrained in both the solutions for TM mixing, there are
no constraints on Majorana-type CP violating phases $\alpha$ and 
$\beta$.

\section{Conclusion}

In conclusion, TM matrix has been defined as a unitary matrix
in which the three elements of the middle column are equal and their 
phases can be rephased away by phase redefinition of lepton fields.
This definition of strong trimaximality is not rephasing invariant and
differs from the weak trimaximality which demands only the equality
of magnitude of the elements of the middle column without putting
any constraints on the phases and, consequently, is less restrictive.
The unitarity constraints on the TM mixing matrix have been solved and
two solutions have been obtained.  The first solution is CP 
conserving with vanishing Jarlskog rephasing invariant $J$ and
has three free parameters \textit{viz.} $ReU_{11}$, $ReU_{13}$ and $ImU_{13}$
which completely determine the mixing matrix. The deviation of 
$\theta_{23}$ from maximality scales up with $\theta_{13}$
and is within experimental constraints. The second solution, called
CP violating solution, has  $ReU_{21}$ as one more free parameter 
and reduces to the CP conserving solution in the limiting case of 
vanishing $J$. The CP conserving solution defines the upper and lower 
bounds within which $\theta_{23}$ varies with respect to $\theta_{13}$ 
in the CP violating solution. So, $J$ vanishes whenever $\theta_{13}$
is zero or $\theta_{23}$ equals its upper or lower bound for the given 
value of $\theta_{13}$. The Jarlskog invariant $J$ takes its 
maximal value $\frac{1}{2\sqrt{3}}|U_{11}||U_{13}|$ when $\theta_{23}$ 
is maximal. So, a notion of maximal CP violation naturally arises 
here: CP violation is maximal for maximal 2-3 mixing. However, the 
Dirac-type CP violating phase $\delta$ does not takes its 
maximum value at the point of the maximal CP violation. The maximal 
value of $\delta$ occurs when $\theta_{23}$ is slightly less than its
maximal value. The allowed range within which $\delta$ can vary
shows a marginal variation of about $10$ degrees with $\theta_{13}$.
For a given value of $\theta_{13}$, 
$s^2_{23}$ and $J$ form an ellipse whose center is defined by 
the TBM mixing and whose axes are governed by the maximal values of  
CP violation and deviation from maximal 2-3 mixing. Trimaximality puts 
no constraints on the Majorana-type CP violating 
phases $\alpha$ and $\beta$ since it is a mass-independent texture.

\newpage
\begin{figure}[t]
\begin{center}
\includegraphics[width=0.5\linewidth]{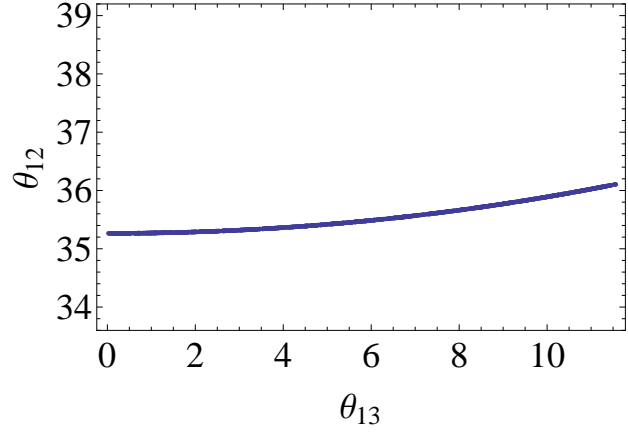}
\end{center}
\caption{Relationship between $\theta_{12}$ and $\theta_{13}$.}
\end{figure}

\begin{figure}[t]
\begin{center}
\includegraphics[width=0.60\linewidth]{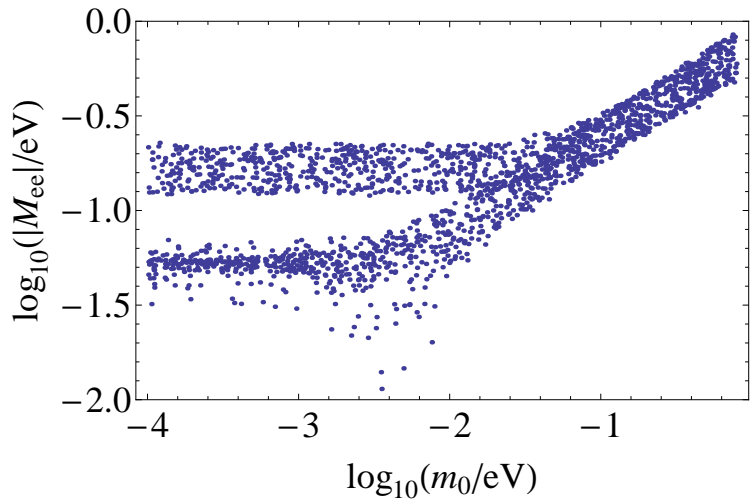}
\end{center}
\caption{The effective Majorana mass $|M_{ee}|$ as
the function of lightest neutrino mass $m_0$ for normal hierarchy
(lower branch) and inverted hierarchy (upper branch).}
\end{figure}

\begin{figure}[t]
\begin{center}
\includegraphics[width=0.60\linewidth]{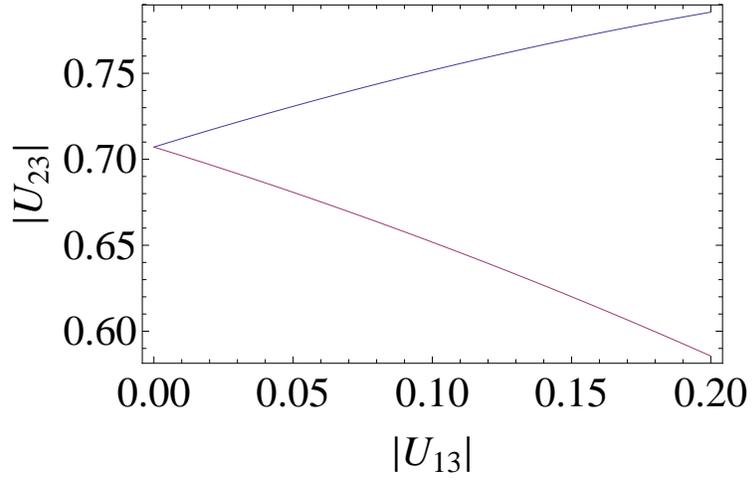}
\end{center}
\caption{$|U_{23}|$ as a function of $|U_{13}|$ for CP conserving solution.}
\end{figure}

\begin{figure}[t]
\begin{center}
\includegraphics[width=0.60\linewidth]{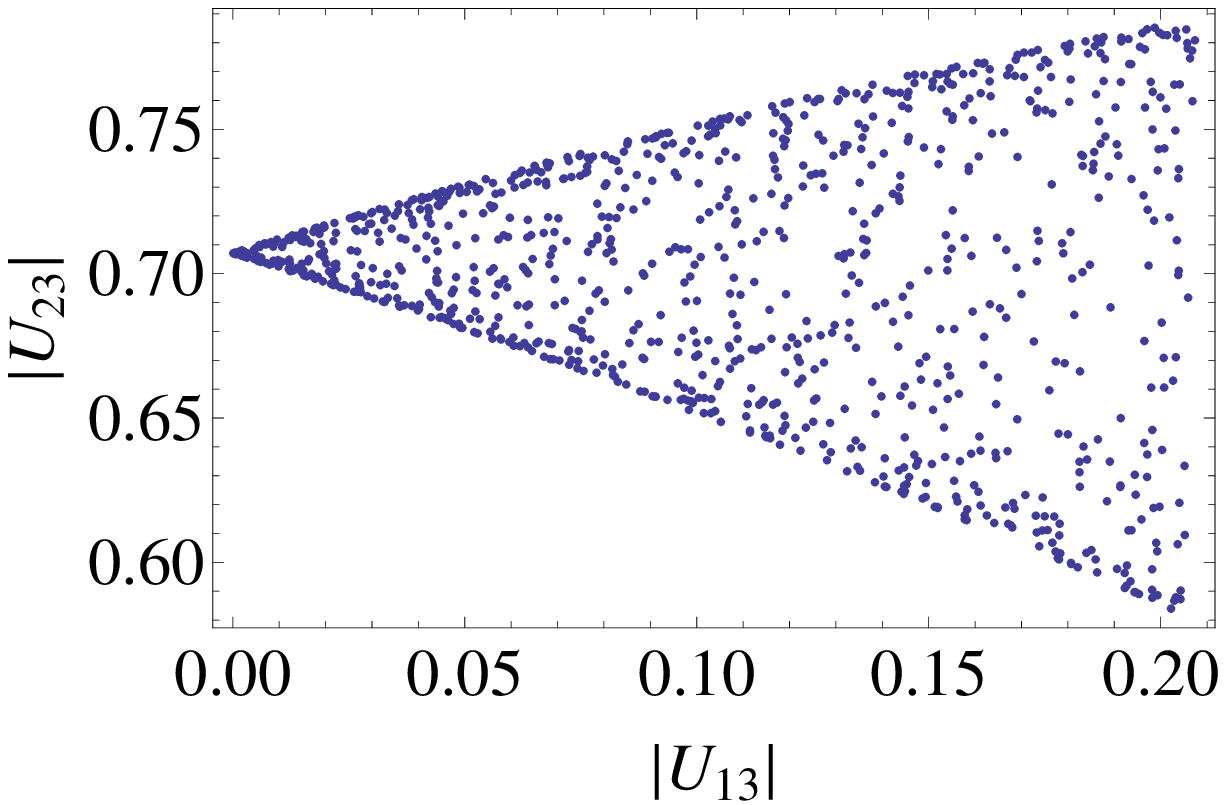}
\end{center}
\caption{$|U_{23}|$ as a function of $|U_{13}|$ for CP vioating solution.}
\end{figure}

\begin{figure}[t]
\begin{center}
\includegraphics[width=0.60\linewidth]{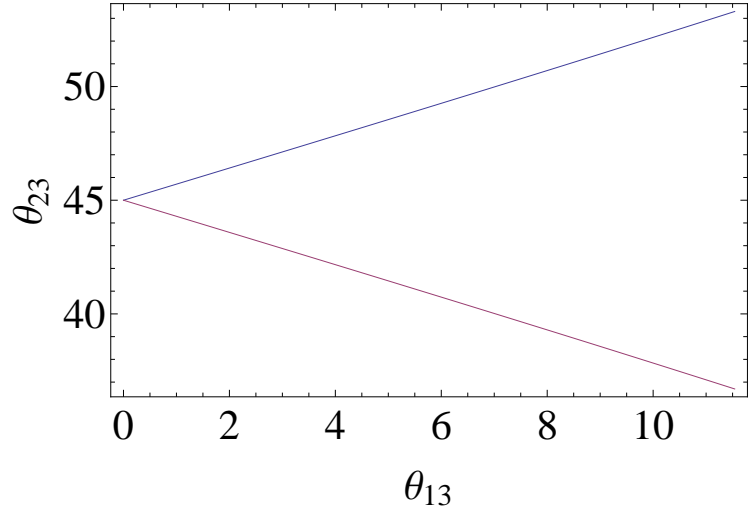}
\end{center}
\caption{Relationship of $\theta_{23}$ with $\theta_{13}$ for CP conserving solution.}
\end{figure}

\begin{figure}[t]
\begin{center}
\includegraphics[width=0.60\linewidth]{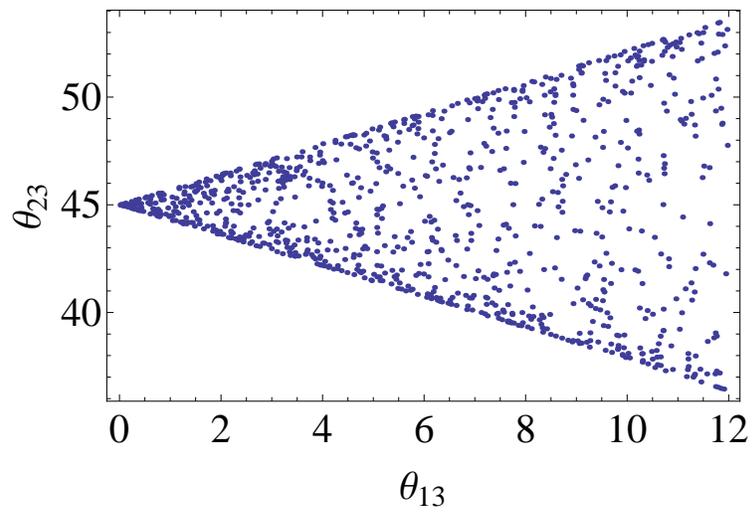}
\end{center}
\caption{Relationship of $\theta_{23}$ with $\theta_{13}$ for CP violating solution.}
\end{figure}

\begin{figure}[t]
\begin{center}
\includegraphics[width=0.60\linewidth]{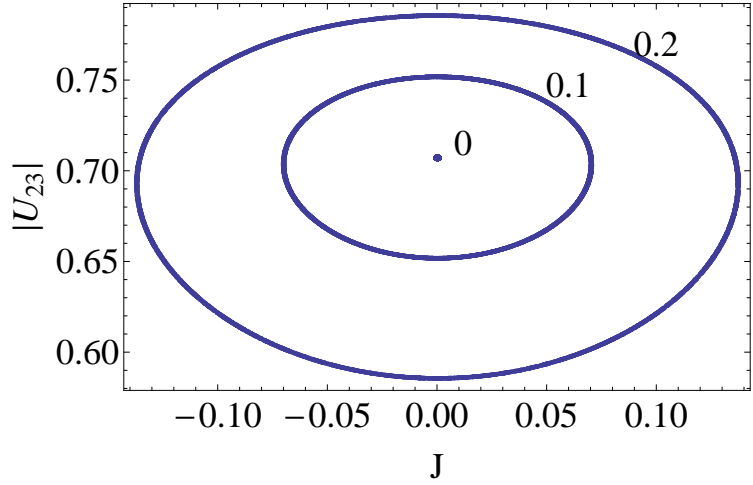}
\end{center}
\caption{Relationship between $J$ and $|U_{23}|$ for $|U_{13}|=0,0.1$ and $0.2$.}
\end{figure}

\begin{figure}[t]
\begin{center}
\includegraphics[width=0.60\linewidth]{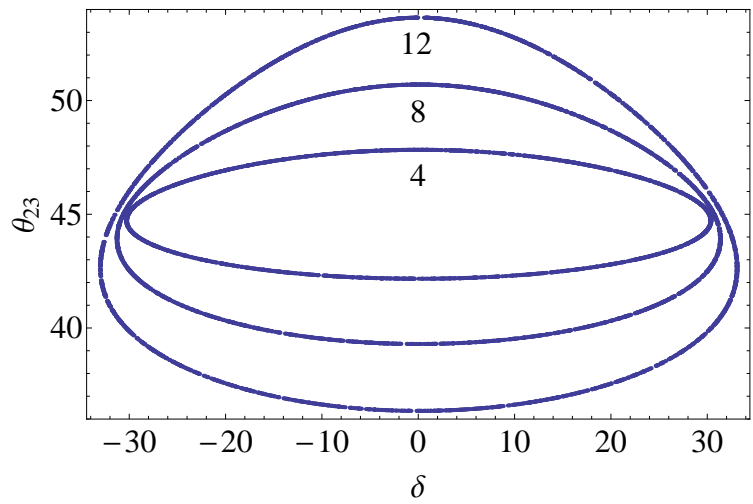}
\end{center}
\caption{Relationship between $\delta$ and $\theta_{23}$ for 
$\theta_{13}=4$, $8$ and $12$ degrees.}
\end{figure}

\begin{figure}[t]
\begin{center}
\includegraphics[width=0.60\linewidth]{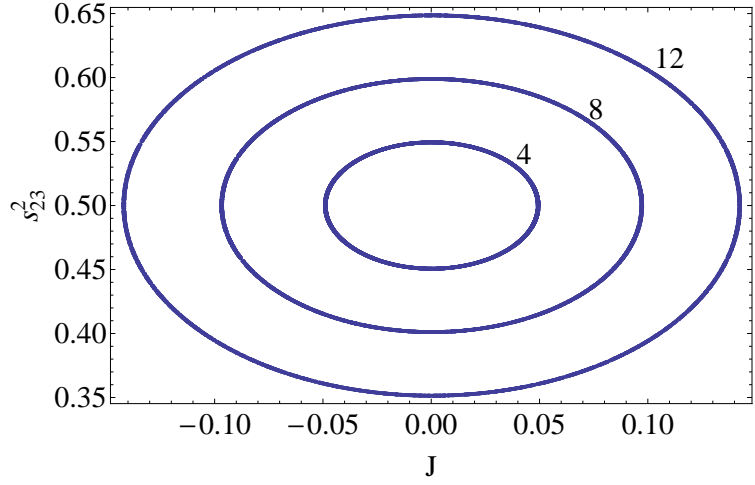}
\end{center}
\caption{The variation of $J$ with $s_{23}^2$ for 
$\theta_{13}=4$, $8$ and $12$ degrees.}
\end{figure}

\begin{figure}[t]
\begin{center}
\includegraphics[width=0.60\linewidth]{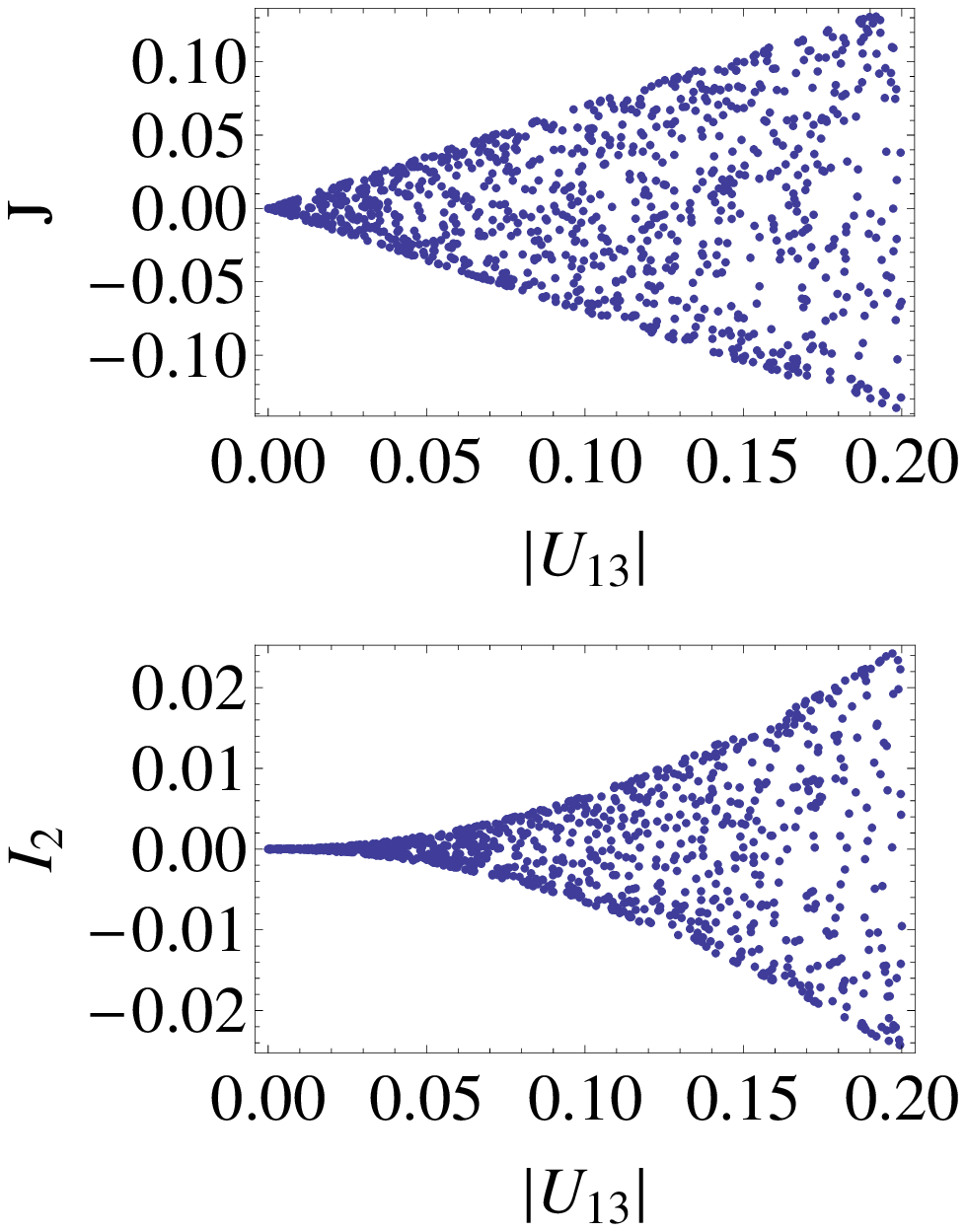}
\end{center}
\caption{The variation of $J$ and $I_2$ with 
$|U_{13}|$.}
\end{figure}


\begin{thebibliography}{}

\bibitem{pdg} C. Amsler et al. [Particle Data Group], 
Phys. Lett. B 667, 1 (2008).

\bibitem{tbm} P. F. Harrison, D. H. Perkins and W. G. Scott, 
Phys. Lett. B530, 167 (2002).

\bibitem{lam-tm-magic} C. S. Lam, Phys. Lett. B 640, 260 (2006).

\bibitem{lam-a4} C. S. Lam, arXiv:hep-ph/0907.2206.

\bibitem{jarlskog} C. Jarlskog, Phys. Rev. Lett. 55, 1039 (1985).

\bibitem{rephasing_inv} Elizabeth Jenkins and Aneesh V. Manohar, 
Nucl. Phys. B 792, 187 (2008).

\bibitem{pdg-para} G. L. Fogli et al., Prog. Part. Nucl. Phys. 57, 742
(2006).

\bibitem{lam-mit} C. S. Lam, Phys. Rev. D 74, 113004 (2006).

\bibitem{tm-albright} C. H. Albright and W. Rodejohann, 
Eur. Phys. J. C 62, 599 (2009).

\bibitem{bhs} James D. Bjorken, P.F. Harrison and W. G. Scott, 
Phys. Rev. D 74, 073012 (2006).

\bibitem {lam-magic} C. S. Lam, Phys. Lett. B 656, 193 (2007).

\bibitem{minimal-a4} Xiao-Gang He and A. Zee, Phys. Lett. B 645, 427 (2007).

\bibitem{tm model} W. Grimus and L. Lavoura, JHEP 0809, 106 (2008).

\bibitem{tm-cp conserving} W. Grimus, L. Lavoura and A. Singraber, Phys. Lett. B 686, 141 (2010).

\end{thebibliography}
\end{document}